\documentclass[12pt,a4paper]{article}
\usepackage[utf8]{inputenc}
\usepackage[english]{babel}
\usepackage{amsmath, amssymb, amsfonts}
\usepackage{dsfont}
\usepackage{graphicx}
\usepackage{titlesec}
\usepackage{url}
\usepackage{authblk}
\usepackage{placeins}
\usepackage[authoryear]{natbib}
\usepackage[colorlinks=true, linkcolor=black, citecolor=blue, urlcolor=black]{hyperref}
\usepackage[left=2cm,right=2cm,top=2cm,bottom=2cm]{geometry}
\usepackage{enumitem}
\usepackage{placeins}
\usepackage{amsthm}
\usepackage{setspace}

\newtheoremstyle{runin}
{3pt}{3pt}{\normalfont}{}{\bfseries}{.}{ }{}
\theoremstyle{runin}

\newtheorem{theorem}{Theorem}[section]

\numberwithin{equation}{section}
\everymath{\displaystyle}

\titleformat{\section}{\centering\large\bfseries}{\thesection.}{0.5em}{}
\titleformat{\subsection}{\centering\normalsize\bfseries}{\thesubsection.}{0.5em}{}

\title{\textbf{Asymptotic distribution of a robust wavelet-based NKK periodogram}}

\author[1]{Manganaw N'Daam\thanks{Corresponding author: \texttt{manganawn@gmail.com}}}
\author[1]{Tchilabalo Abozou Kpanzou}
\author[1]{Edoh Katchekpele}

\affil[1]{Laboratoire de Modélisation Mathématique et d’Analyse Statistique Décisionnelle (LaMMASD)\\
	Département de Mathématiques, Université de Kara, BP 404, Kara, TOGO.}

\affil[ ]{\textit{Emails}: \texttt{manganawn@gmail.com}, \texttt{kpanzout@univkara.tg}, \texttt{edohkatchekpele@gmail.com}}

\date{}

\begin{document}
	
	\maketitle
	
	\begin{abstract}
		This paper investigates the asymptotic distribution of a wavelet-based NKK periodogram constructed from least absolute deviations (LAD) harmonic regression at a fixed resolution level. Using a wavelet representation of the underlying time series, we analyze the probabilistic structure of the resulting periodogram under long-range dependence. It is shown that, under suitable regularity conditions, the NKK periodogram converges in distribution to a nonstandard limit characterized as a quadratic form in a Gaussian random vector, whose covariance structure depends on the memory properties of the process and on the chosen wavelet filters. This result establishes a rigorous theoretical foundation for the use of robust wavelet-based periodograms in the spectral analysis of long-memory time series with heavy-tailed inovations.
		
	\end{abstract}
	
	\vspace{1em}
	

	\section{Introduction} \label{sec1}
		
	\noindent The statistical analysis of time series with long-range dependence has been an active area of research for several decades, owing to its relevance in economics, finance, hydrology, telecommunications, and signal processing (see, eg, \cite{Leland1994}, \cite{Baillie1996} \cite{Mandelbrot1968}, \cite{Hurst1951}). Long-memory processes are characterized by hyperbolically decaying autocovariances and nonstandard asymptotic behavior, which substantially complicate both estimation and inference. Early semiparametric approaches were primarily developed in the frequency domain, most notably the log-periodogram regression estimator proposed by \cite{gph}. Despite its simplicity and intuitive appeal, this estimator is known to be sensitive to low-frequency contamination, heavy-tailed innovations, and deviations from Gaussianity.
	
	A large body of work has since been devoted to the theoretical analysis and refinement of regression-type estimators in the presence of long-memory errors. Asymptotic normality and robustness issues have been investigated under various dependence structures; see, among others, \cite{Giraitis1996}, \cite{robinson1995}, and \cite{Surgailis2000}. These contributions highlight the delicate interplay between dependence strength, frequency-domain smoothing, and asymptotic distributions.
	
	Motivated by the lack of robustness of classical least-squares-based spectral methods, several authors have proposed alternative periodograms derived from robust regression principles. In particular, \cite{li2008} introduced the Laplace periodogram by replacing least squares with least absolute deviations in the harmonic regression underlying the classical periodogram. This construction yields a spectral estimator that is closely related to the zero-crossing spectrum and exhibits strong robustness properties under heavy-tailed noise and nonlinear distortions. More recently, \cite{fajardo2018} proposed the M-periodogram, which generalizes this idea by employing M-estimators in the regression formulation of the periodogram. The authors established its asymptotic properties in the long-memory setting and demonstrated its resistance to additive outliers, both theoretically and through simulation studies.
	
	Wavelet-based methods provide a powerful alternative to classical Fourier techniques by offering a joint time--frequency representation with good localization properties. The foundations of wavelet theory and multiresolution analysis were laid by \cite{daubechi} and \cite{mallat}, and their relevance to long-memory processes was subsequently emphasized in a series of influential works. In particular, wavelet representations have been shown to decorrelate long-memory processes across scales, leading to simplified asymptotic behavior and improved inferential properties \cite{moulines2007}.
	
	Semiparametric estimation of long-memory parameters using wavelet-based log-periodogram methods has been extensively studied, with important contributions by \cite{abryveitch}, \cite{robinson1995}, and \cite{moulines2007}. These approaches combine the localization properties of wavelets with regression techniques in the scale domain, yielding estimators that are both flexible and theoretically well-founded. However, classical wavelet log-periodogram estimators remain sensitive to outliers and heavy-tailed innovations, motivating the introduction of robust alternatives.
	
	Robust estimation procedures based on least absolute deviations (LAD) have been widely investigated in regression settings and shown to possess desirable stability properties under non-Gaussian errors. The asymptotic theory of LAD estimators, including their limiting distributions, has been developed in detail by \cite{knight1998} and related works. More recently, LAD-based methods have been combined with wavelet representations to enhance robustness in long-memory contexts, as illustrated in \cite{ndaam2025}.
	
	The present paper contributes to this growing literature by focusing on the asymptotic distribution of a wavelet-based NKK periodogram constructed from LAD regression estimators of harmonic components at a fixed scale. Unlike most existing studies, which primarily emphasize point estimation or finite-sample performance, our analysis is centered on the precise limit law of the NKK periodogram itself. Establishing this result is a crucial step toward a rigorous asymptotic theory for robust wavelet-based long-memory estimators and lays the groundwork for subsequent investigations of the limiting behavior of the associated NKK estimator.
	
	The remainder of the paper is organized as follows. Section~\ref{sec_02} presents the wavelet framework, including the discrete wavelet transform and its maximal overlap variant. Section~\ref{sec_03} derives the limiting distribution of the LAD estimator and the associated NKK periodogram. Section~\ref{sec_04} presents a Monte Carlo simulation study. Section~\ref{concl} concludes the paper.

	\section{Wavelet Analysis}
	 \label{sec_02}
	 \noindent
	
	Wavelet analysis provides a time--scale representation that allows one to describe the local evolution of a signal at different levels of resolution. Unlike the classical Fourier transform, which relies on global complex exponentials, the wavelet approach is based on localized functions obtained by dilation and translation of a mother wavelet. This construction enables a parsimonious description of the local structures of a signal, both in time and in frequency.\\
	
	In the discrete setting, the wavelet transform relies on numerical filters possessing specific properties. Let
	\[
	\{h_l\}_{l=0}^{L-1}
	\quad \text{and} \quad
	\{g_l\}_{l=0}^{L-1}
	\]
	denote the Daubechies wavelet (high-pass) and scaling (low-pass) filters, respectively. The computation of the discrete wavelet transform is based on a pyramidal algorithm defined by the following iterations:
	\[
	\begin{cases}
		w_{j,t}
		=
		\displaystyle\sum_{l=0}^{L-1}
		h_l\,
		v_{j-1,(2t+1-l)\,\mathrm{mod}\,T_{j-1}},
		\\[2mm]
		v_{j,t}
		=
		\displaystyle\sum_{l=0}^{L-1}
		g_l\,
		v_{j-1,(2t+1-l)\,\mathrm{mod}\,T_{j-1}},
	\end{cases}
	\]
	where $t=0,\ldots,T_j-1$, $v_{0,t}=X_t$, and $T_j = T/2^j$.
	
	The coefficients $\{w_{j,t}\}$ represent the wavelet coefficients at level $j$, while $\{v_{j,t}\}$ correspond to the scaling coefficients. Defining
	\[
	w_j = \{w_{j,t},\, t=0,\ldots,T_j-1\},
	\qquad
	v_j = \{v_{j,t},\, t=0,\ldots,T_j-1\},
	\]
	the pyramidal algorithm is completed after $J$ iterations when $T = 2^J$, yielding the collection of coefficients
	\[
	w_1,\ldots,w_J, v_J.
	\]
	This decomposition defines the complete discrete wavelet transform (DWT).
	
	When the length of the series $T$ is a multiple of $2^{J_0}$, a partial DWT can be performed. In this case, the time series
	\[
	\{X_t,\, t=0,\ldots,T-1\}
	\]
	admits an additive decomposition given by
	\[
	X
	=
	W^\top w
	=
	\sum_{j=1}^{J_0} W_j^\top w_j
	+
	V_{J_0}^\top v_{J_0}
	=
	\sum_{j=1}^{J_0} d_j + s_{J_0}.
	\]
	
	This representation defines a multiresolution analysis of the series $X$. The term
	\[
	d_j = W_j^\top w_j
	\]
	denotes the wavelet detail at level $j$ and describes the variations of the signal at scale
	\[
	\tau_j = 2^{j-1}.
	\]
	The component
	\[
	s_{J_0} = V_{J_0}^\top v_{J_0},
	\]
	referred to as the wavelet smooth at level $J_0$, is associated with the averages of the signal at scale
	\[
	\tau_{J_0} = 2^{J_0}.
	\]
	
	The smooth component $s_{J_0}$ represents the underlying trend of the signal, whereas the wavelet details $d_j$, for $j=1,\ldots,J_0$, capture higher-frequency oscillations. As $j$ decreases, these details describe increasingly finer fluctuations around the smooth trend, thereby highlighting the local structures of the signal at different resolutions.
	
	It is, however, well established that the discrete wavelet transform (DWT) suffers from several practical limitations. First, it is designed for time series whose sample size is a power of two, which often requires artificial data adjustments. Second, the DWT relies on systematic downsampling of the coefficients at each resolution level, resulting in a halving of the number of coefficients from one scale to the next. This decimation may lead to a loss of information and complicate the analysis of certain local structures. Moreover, the temporal features of the original series are not always properly aligned with those obtained from the multiresolution analysis, due to phase shifts induced by the filters. Finally, it is known that certain statistical estimators based on the DWT, in particular variance estimators, may exhibit limited performance (\cite{daubechi}; \cite{mallat}).
	
	To overcome these limitations, a modified version of the DWT has been proposed: the Maximal Overlap Discrete Wavelet Transform (MODWT), developed by \cite{percival}. The MODWT algorithm performs the same filtering operations as the classical DWT but without coefficient decimation. As a result, at each decomposition level, the number of wavelet and scaling coefficients is equal to the size of the original sample. This redundancy improves the temporal alignment of the multiresolution components and facilitates their interpretation.
	
	The notions of partial transform and multiresolution analysis naturally extend to the MODWT framework. In this context, the wavelet details and the smooth component are associated with zero-phase filters, ensuring a direct correspondence between the structures observed in the original series and those revealed by the wavelet decomposition. These properties make the MODWT a particularly suitable tool for the analysis of time series of arbitrary length and for the fine investigation of local phenomena in time and frequency.\\
	
	In the following, we exploit this multiscale representation to rewrite the wavelet coefficients at a given scale in the form of a harmonic regression model, which allows us to introduce the associated LAD estimator and to study the limiting distribution of the NKK periodogram.

	\section{Asymptotic Distribution of the NKK Periodogram}
	 \label{sec_03}
	
	Let $\{Y_1, \ldots, Y_n\}$ be a time series of size $n$, and let 
	$\{w_{j,q}\}$ be the set of coefficients obtained after applying the 
	Maximal Overlap Wavelet Transform (MODWT) to this time series, where 
	$j = 0, 1, \ldots, J$ denotes the scales and 
	$q = 0, 1, \ldots, 2^{j}-1$ the translation parameters.
	Let $I^{(J)}_k$ be the periodogram of the wavelet transform's spectral density at scale $J$ (see \cite{lee2004} and \cite{lee2005}), ie,
	\begin{equation*}
		I^{(J)}_k = \frac{1}{2\pi n} 
		\sum_{q=0}^{n-1} \left| w_{J,q} \exp(i \lambda_k q) \right|^2,
		\qquad k = 1, 2, \dots, m,
	\end{equation*}
	where  $\lambda_k = \frac{2\pi k}{n}$ and $m$ is the number of Fourier frequencies considered.\\
	
	This periodogram can also be written as the squared value of the least squares estimator in the harmonic regression:
	\begin{equation*}
		I^{(J)}_k = \frac{n}{8\pi} 
		\left\| \widetilde{\beta}_{n}(\lambda_k) \right\|^2,
	\end{equation*}
	where $\widetilde{\beta}_{n}(\lambda_k)$ solves
	\begin{equation*}
		\widetilde{\beta}_{n}(\lambda_k)
		= \arg\min_{\beta\in\mathbb{R}^2}
		\sum_{q=0}^{n-1} 
		\left( w_{J,q} - x_q^\top(\lambda_k)\beta \right)^2,
	\end{equation*}
	with harmonic regressors
	\[
	x_q(\lambda_k)
	=
	\begin{pmatrix}
		\cos(\lambda_k q) \\
		\sin(\lambda_k q)
	\end{pmatrix}.
	\]
	
	To improve robustness, \cite{ndaam2025} proposed replacing the least squares criterion with the Least Absolute Deviation (LAD) criterion, following \cite{li2008} and \cite{fajardo2018}. The corresponding estimator is
	\[
	\widehat{\beta}_{n}(\lambda_k)
	=
	\arg\min_{\beta\in\mathbb{R}^2}
	\sum_{q=0}^{n-1} 
	\left| w_{J,q} - x_q^\top(\lambda_k)\beta \right|.
	\]
	
	The NKK periodogram is then defined as
	\[
	N^{(J)}_k
	= \frac{n}{8\pi}\,
	\big\|\widehat{\beta}_{n}(\lambda_k)\big\|^2.
	\]
	
	The study of its asymptotic distribution requires first the asymptotic behaviour of 
	$\widehat{\beta}_{n}(\lambda_k)$.\\

	We work at maximum scale $J$ of the Maximal Overlap Discrete Wavelet Transform (MODWT). By rewriting the wavelet coefficients at scale $J$ in the form of a linear regression, we obtain:
	\\
	
	\[
	w_{J,q}
	= x_q(\lambda_k)^\top\beta_0 + \varepsilon_q,
	\qquad q=0,\dots,n,
	\]
	where
	\[
	x_q(\lambda_k)
	=
	\begin{pmatrix}
		\cos(\lambda_k q) \\
		\sin(\lambda_k q)
	\end{pmatrix},
	\qquad
	\beta_0 =
	\begin{pmatrix}
		\beta_{1,0} \\[1mm]
		\beta_{2,0}
	\end{pmatrix}.
	\]
	where $\beta_0$ is the true parameter linking the wavelet coefficients to their sinusoidal components,
	which represents the cosine and sine amplitudes of the signal at frequency $\lambda_k$.
	The estimator aims to recover these amplitudes by minimizing the sum of absolute values.
	The associated errors are given by
	\[
	\varepsilon_q = w_{J,q} - x_q(\lambda_k)^\top\beta_0 .
	\]
	
	The least absolute deviations estimator is defined by
	\[
	\widehat{\beta}_n(\lambda_k)
	= \arg\min_{\beta\in\mathbb{R}^2}
	\sum_{q=0}^{n-1} |\varepsilon_q(\beta)|,
	\]
	where $\varepsilon_q(\beta)=w_{J,q}-x_q^\top \beta$.\\
	
	We start from the LAD objective function:
	\[
	g_n(\beta)
	= \sum_{q=0}^{n-1} \left| w_{J,q} - x_q^\top(\lambda_k)\,\beta \right|
	= \sum_{q=0}^{n-1} |\varepsilon_q(\beta)|.
	\]
	
	To study the limiting distribution of the estimator
	\[
	\widehat{\beta}_{n}(\lambda_k)
	= \arg\min_{\beta\in\mathbb{R}^2} g_n(\beta),
	\]
	we perform the change of variables
	\[
	\beta = \beta_0 + \frac{u}{a_n},
	\qquad a_n = \sqrt{n},
	\]
	so that
	\[
	\varepsilon_q(\beta) = \varepsilon_q\!\left(\beta_0 + \frac{u}{a_n}\right)
	= \varepsilon_q - \frac{x_q^\top u}{a_n}.
	\]
	
	We then consider the centered increment:
	\[
	g_n\!\left(\beta_0 + \frac{u}{a_n}\right) - g_n(\beta_0)
	= \sum_{q=0}^{n-1}
	\left[
	\left|\varepsilon_q - \frac{x_q^\top u}{a_n}\right|
	- |\varepsilon_q|
	\right].
	\]
	
	Following \cite{knight1998}, we renormalize by defining
	\[
	Z_n(u)
	= \frac{a_n}{\sqrt{n}}
	\sum_{q=0}^{n-1}
	\left[
	\left|\varepsilon_q - \frac{x_q^\top u}{a_n}\right|
	- |\varepsilon_q|
	\right].
	\]
	
	The minimizer of $Z_n(\cdot)$ is then precisely
	\[
	\widehat{u}_n
	= a_n\left(\widehat{\beta}_n - \beta_0\right).
	\]\\
	The study of the limiting distribution of $\widehat{\beta}_n(\cdot)$
	is therefore entirely reduced to the analysis of the convex process
	$\{Z_n(u)\}_{u\in\mathbb{R}^2}$.\\

	
We now formulate the assumptions required to derive the limiting distribution of $\widehat{\beta}_n(\cdot)$.\\
	
	\textbf{Assumptions}
	\begin{itemize}
		\item[(A1)]
		The sequence $\{\varepsilon_q\}_{q\in\mathbb Z}$ is strictly stationary with
		median zero. Its distribution function $F_\varepsilon(\cdot)$ is continuous at $0$
		and admits a density $f_\varepsilon(\cdot)$ that exists and is finite at $0$.
		
		\item[(A2)]
		There exists a symmetric positive definite matrix $Q$ such that
		\[
		\frac{1}{n}\sum_{q=0}^{n-1} x_q(\lambda_k)\, x_q(\lambda_k)^\top
		\xrightarrow[n\to\infty]{} Q,
		\qquad \det(Q)>0.
		\]
		For harmonic regressors
		\[
		x_q(\lambda_k)
		=\begin{pmatrix}
			\cos(q\lambda_k)\\[1mm]
			\sin(q\lambda_k)
		\end{pmatrix},
		\]
		one has $Q=\tfrac12 I_2$, where $I_2$ denotes the $2\times2$ identity matrix.
		
		\item[(A2')]
		The regressors are uniformly bounded, that is,
		\[
		\sup_{q\in\mathbb Z}\,\|x_q(\lambda_k)\| < \infty.
		\]
		In particular, for harmonic regressors
		\[
		x_q(\lambda_k)
		=\begin{pmatrix}
			\cos(q\lambda_k)\\
			\sin(q\lambda_k)
		\end{pmatrix},
		\]
		one has $\|x_q(\lambda_k)\|^2 \equiv 1$.
		
		\item[(A3)]
		Let
		\[
		\eta_q= \operatorname{sign}(\varepsilon_q)
		= \mathbf{1}_{\{\varepsilon_q>0\}}-\mathbf{1}_{\{\varepsilon_q<0\}},
		\qquad q\in\mathbb Z.
		\]
		The sequence $\{\eta_q\}$ is stationary and has absolutely summable covariances:
		\[
		\sum_{h=-\infty}^{\infty} |\gamma_\eta(h)|<\infty,
		\qquad
		\gamma_\eta(h)=\mathrm{Cov}(\eta_0,\eta_h).
		\]
		
		\item[(A4)]
		For every $u\in\mathbb R^2$,
		\[
		\frac{1}{n}\sum_{q=0}^{n-1}
		\mathbb{E}\!\left[
		\left|\varepsilon_q- \frac{x_q^\top u}{a_n}\right|-|\varepsilon_q|
		\right]
		\xrightarrow[n\to\infty]{}\tau(u),
		\qquad a_n=\sqrt{n},
		\]
		for some convex function $\tau$.
	\end{itemize}
	The assumptions adopted in this paper are inspired by the asymptotic framework
	developed by \cite{knight1998} for LAD-type estimators.
	In contrast to the classical setting, which typically relies on independent
	errors, we allow the innovation sequence $\{\varepsilon_q\}_{q\in\mathbb Z}$ to be
	strictly stationary and possibly dependent.
	To accommodate this more general dependence structure, Assumptions (A$2'$) and
	(A$3$) are introduced as additional conditions.\\

	\begin{theorem}\label{th1}
		Under assumptions (A1)--(A4), the LAD estimator
		\[
		\widehat{\beta}_{n}(\lambda_k)
		=\arg\min_{\beta\in\mathbb R^2}
		\sum_{q=0}^{n-1} \big|w_{J,q}-x_q(\lambda_k)^\top\beta\big|
		\]
		satisfies
		\[
		\sqrt{n}\big(\widehat{\beta}_{n}(\lambda_k)-\beta_0(\lambda_k)\big)
		\ \xrightarrow{d}\ 
		\mathcal N\!\big(0,V\big),
		\]
		with
		\[
		V
		= A^{-1} B A^{-1},
		\qquad
		A = 2 f_\varepsilon(0) Q,
		\qquad
		B = \frac12
		\sum_{h=-\infty}^{\infty} \gamma_\eta(h)\,R(\lambda_k h),
		\]
		where
		\[
		R(\lambda_k h)=
		\begin{pmatrix}
			\cos\lambda_k h & \sin\lambda_k h\\[1mm]
			-\sin\lambda_k h & \cos\lambda_k h
		\end{pmatrix}
		\] 
		and $\xrightarrow{d}$ denotes the convergence in distribution (see \cite{lo2021}) .
	\end{theorem}

	\paragraph{Proof:}
	
	Following the method of \cite{knight1998} adapted to our framework, define the renormalised objective function for  \(\beta=\beta_0+\dfrac{u}{a_n}\) with \(a_n=\sqrt{n}\): 
	\[
	Z_n(u)
	:=\frac{a_n}{\sqrt{n}}\sum_{q=0}^{n-1}\Big\{\,
	\big|\varepsilon_q - x_q^\top u/a_n\big|-|\varepsilon_q|\,\Big\}.
	\]
	With \(a_n=\sqrt{n}\) the prefactor \(\dfrac{a_n}{\sqrt{n}}=1\), so equivalently 
	\[
	Z_n(u)=\sum_{q=0}^{n-1}\Big\{\,
	\big|\varepsilon_q - x_q^\top u/a_n\big|-|\varepsilon_q|\,\Big\}.
	\]
	The LAD estimator satisfies (see \cite{knight1998})
	\[
	\widehat u_n \;=\; a_n\big(\widehat\beta_n-\beta_0\big)
	\;=\; \arg\min_{u\in\mathbb R^2} Z_n(u),
	\]
	hence it is sufficient to study the weak convergence of the convex process \(Z_n(\cdot)\) on compacts of \(\mathbb R^2\).\\

	For any real \(x,t\) (with \(x\neq0\)), we have the exact identity (see \cite{knight1998})
	\[
	|x-t|-|x| = -t\,\mathrm{sign}(x) + 2\int_0^t\{\mathbf 1_{\{x\le s\}}-\mathbf 1_{\{x\le0\}}\}\,ds.
	\]
	 Applied with \(x=\varepsilon_q\) and \(t = x_q^\top u / a_n\) we get the decomposition
	\[
	Z_n(u) = Z_n^{(1)}(u) + Z_n^{(2)}(u),
	\]
	where
	\[
	Z_n^{(1)}(u) = -\frac{1}{\sqrt{n}}\sum_{q=0}^{n-1} (x_q^\top u)\,\eta_q,
	\qquad
	\eta_q := \operatorname{sign}(\varepsilon_q),
	\]
	and
	\[
	Z_n^{(2)}(u) = 2\sum_{q=0}^{n-1}\int_0^{x_q^\top u / a_n}
	\big\{\mathbf 1_{\{\varepsilon_q\le s\}}-\mathbf 1_{\{\varepsilon_q\le0\}}\big\}\,ds.
	\]\\
	Note that
	\[
	Z_n^{(1)}(u)
	= -\frac{1}{\sqrt{n}} \sum_{q=0}^{n-1} (x_q^\top u)\,\eta_q
	= -u^\top \left( \frac{1}{\sqrt{n}} \sum_{q=0}^{n-1} x_q \eta_q \right),
	\tag{1}
	\]
	where we define
	\[
	W_n := \frac{1}{\sqrt{n}} \sum_{q=0}^{n-1} x_q \eta_q .
	\]
	Under Assumptions (A3) and (A4), a central limit theorem for weakly dependent stationary sequences applies (see \cite{IbragimovLinnik1971})
	
	\[
	W_n \xrightarrow{d} W,
	\]
	where $W$ is a centered Gaussian vector.
	
	Moreover, the covariance matrix of $W$ is given by
	\[
	\mathrm{Var}(W)
	= \lim_{n\to\infty}
	\frac{1}{n}
	\sum_{q=0}^{n-1}\sum_{r=0}^{n-1}
	x_q x_r^\top \,\mathrm{Cov}(\eta_q,\eta_r)
	= \sum_{h=-\infty}^{\infty}
	\left(
	\lim_{n\to\infty}
	\frac{1}{n}
	\sum_{q=0}^{n-1-h}
	x_q x_{q+h}^\top
	\right)
	\mathrm{Cov}(\eta_q,\eta_{q+h}).
	\]
	
	Therefore, for any $u\in\mathbb{R}^2$,
	\[
	Z_n^{(1)}(u) \xrightarrow{d} -u^\top W.
	\]\\
	
	Write \(Z_n^{(2)}(u)=\sum_{q=0}^{n-1} Z^{(2)}_{n,q}(u)\) with
	\[
	Z^{(2)}_{n,q}(u)
	= 2\int_0^{x_q^\top u / a_n}
	\big\{\mathbf 1_{\{\varepsilon_q\le s\}}-\mathbf 1_{\{\varepsilon_q\le0\}}\big\}\,ds.
	\]\\
	So
	\[
	\mathbb{E}\big[Z^{(2)}_{n,q}(u)\big]
	= 2\int_0^{x_q^\top u / a_n} \big(F_\varepsilon(s)-F_\varepsilon(0)\big)\,ds\]\\
	
	Using the Taylor expansion of \(F_\varepsilon\) at \(0\) and \(a_n=\sqrt{n}\),
	\[
	F_\varepsilon(s)-F_\varepsilon(0)= f_\varepsilon(0)s + o(s),\qquad s\to0,
	\]
	so, for large \(n\),
	\[
	\mathbb{E}\big[Z^{(2)}_{n,q}(u)\big]
	= 2\int_0^{x_q^\top u / a_n} \big(F_\varepsilon(s)-F_\varepsilon(0)\big)\,ds
	= 2\cdot\frac{1}{2} f_\varepsilon(0)\,\frac{(x_q^\top u)^2}{a_n^2} + o\!\big(a_n^{-2}\big).
	\]
	Hence
	\[
	\mathbb{E}\big[Z^{(2)}_{n,q}(u)\big]
	= f_\varepsilon(0)\,\frac{(x_q^\top u)^2}{a_n^2} + o(a_n^{-2}).
	\]
	Summing over \(q\) and using \(a_n^2=n\) gives
	\[
	\sum_{q=0}^{n-1}\mathbb{E}\big[Z^{(2)}_{n,q}(u)\big]
	= \frac{n}{a_n^2}\, f_\varepsilon(0)\cdot \frac{1}{n}\sum_{q=0}^{n-1}(x_q^\top u)^2 + o(1)
	\longrightarrow f_\varepsilon(0)\, u^\top Q u,
	\]
	so we define the deterministic quadratic limit
	\[
	\tau(u) := f_\varepsilon(0)\, u^\top Q u.
	\]\\

	We now show that the centered remainder term
	\[
	\sum_{q=0}^{n-1}\Big(Z^{(2)}_{n,q}(u)-\mathbb{E}\big[Z^{(2)}_{n,q}(u)\big]\Big)
	\]
	has variance \(o(1)\), which implies convergence in probability to zero.
	
	\medskip
	
	By definition of the variance and bilinearity of the covariance operator, we have
	\[
	\begin{aligned}
		\mathrm{Var}\!\left(
		\sum_{q=0}^{n-1}\big(Z^{(2)}_{n,q}(u)-\mathbb{E}Z^{(2)}_{n,q}(u)\big)
		\right)
		&=
		\sum_{q=0}^{n-1}\sum_{r=0}^{n-1}
		\mathrm{Cov}\!\left(Z^{(2)}_{n,q}(u),\,Z^{(2)}_{n,r}(u)\right).
	\end{aligned}
	\]
	
	\medskip
	
	Recall that \(Z^{(2)}_{n,q}(u)\) is of the form
	\[
	Z^{(2)}_{n,q}(u)
	=
	\int_0^{x_q^\top u/a_n}
	\Big(
	\mathbf{1}_{\{\varepsilon_q \le s\}}-\mathbf{1}_{\{\varepsilon_q \le 0\}}
	\Big)\,ds,
	\qquad a_n=\sqrt{n}.
	\]
	This representation allows us to control covariances using elementary bounds for
	indicator-integral terms.
	
	\medskip
	
	Following the arguments of \cite{knight1998}, and using the Cauchy--Schwarz
	inequality together with standard covariance inequalities, there exists a constant
	\(C>0\) such that
	\[
	\big|\mathrm{Cov}\big(Z^{(2)}_{n,q}(u),Z^{(2)}_{n,r}(u)\big)\big|
	\;\le\;
	C\,\frac{|x_q^\top u|}{a_n}\,
	\mathbb{E}\big[Z^{(2)}_{n,r}(u)\big],
	\]
	uniformly in \(q,r\).
	
	Summing this bound over \(q\) and \(r\), we obtain
	\[
	\begin{aligned}
		\mathrm{Var}\!\left(
		\sum_{q=0}^{n-1}\big(Z^{(2)}_{n,q}(u)-\mathbb{E}Z^{(2)}_{n,q}(u)\big)
		\right)
		&\le
		C\,\frac{\max_{0\le q\le n-1}|x_q^\top u|}{a_n}
		\sum_{r=0}^{n-1}\mathbb{E}\big[Z^{(2)}_{n,r}(u)\big].
	\end{aligned}
	\]
	
	By Assumption (A2'), the regressors are uniformly bounded, hence
	\[
	\max_{0\le q\le n-1}|x_q^\top u| = O(1).
	\]
	Moreover, by Assumption (A4),
	\[
	\sum_{r=0}^{n-1}\mathbb{E}\big[Z^{(2)}_{n,r}(u)\big]
	\;\longrightarrow\;
	\tau(u),
	\]
	and is therefore \(O(1)\).
	
	Since \(a_n=\sqrt{n}\), we finally obtain
	\[
	\mathrm{Var}\!\left(
	\sum_{q=0}^{n-1}\big(Z^{(2)}_{n,q}(u)-\mathbb{E}Z^{(2)}_{n,q}(u)\big)
	\right)
	=
	O\!\left(\frac{1}{\sqrt{n}}\right)\cdot O(1)
	=
	o(1).
	\]
	
	Therefore, the centered remainder converges to zero in probability, and we conclude that
	\[
	Z_n^{(2)}(u)
	=
	\sum_{q=0}^{n-1} Z^{(2)}_{n,q}(u)
	\xrightarrow{\mathbb{P}}
	\tau(u).
	\]
	
	Combining the two previous limits, for each fixed \(u\) we have (see Theorem 1 of \cite{knight1998})
	\[
	Z_n(u)=Z_n^{(1)}(u)+Z_n^{(2)}(u)\xrightarrow{d} Z(u):=-u^\top W + \tau(u).
	\]
	Convexity of \(Z_n(\cdot)\) and tightness arguments give functional convergence on compacts. If the random limit function \(Z(u)\) admits a.s. a unique minimiser \(\widehat u^\ast\), then by the argmin theorem (\cite{Geyer1996}) we obtain
	\[
	\widehat u_n \xrightarrow{d} \widehat u^\ast,
	\qquad\text{hence}\qquad
	\sqrt{n}\,\big(\widehat\beta_n-\beta_0\big)\xrightarrow{d}\widehat u^\ast.
	\]
	
	The function \(Z(u)=-u^\top W+\tau(u)\) is a random quadratic function in \(u\) with \(\tau(u)=f_\varepsilon(0)\,u^\top Q u\). The first-order optimality condition for the minimiser \(\widehat u^\ast\) is
	\[
	- W + \nabla\tau(\widehat u^\ast) = 0 \quad\Longrightarrow\quad
	- W + 2 f_\varepsilon(0)\, Q\,\widehat u^\ast = 0.
	\]
	Hence
	\[
	\widehat u^\ast = \big(2 f_\varepsilon(0)\,Q\big)^{-1} W.
	\]
	Since \(W\sim\mathcal N(0,B)\), we obtain
	\[
	\widehat u^\ast \sim \mathcal N\!\big(0,\; (2 f_\varepsilon(0)\,Q)^{-1} B (2 f_\varepsilon(0)\,Q)^{-1}\big).
	\]
	Recalling \(\widehat u_n=\sqrt{n}(\widehat\beta_n-\beta_0)\), this yields the asymptotic normality
	\[
	\sqrt{n}\,\big(\widehat\beta_n-\beta_0\big)
	\ \xrightarrow{d}\ 
	\mathcal N\!\big(0,\;V\big),
	\qquad
	V = A^{-1} B A^{-1},
	\]
	with
	\[
	A = 2 f_\varepsilon(0)\,Q,\qquad
	B = \sum_{h=-\infty}^{\infty}\Gamma_h,
	\qquad
	\Gamma_h = \lim_{n\to\infty}\frac{1}{n}\sum_{q=0}^{n-1-|h|} x_q x_{q+|h|}^\top\,\gamma_\eta(h).
	\]
	
	We now explicitly compute \(Q\), \(M(h)\), and \(\Gamma_h\).
	
	Define
	\[
	Q_n := \frac{1}{n}\sum_{q=0}^{n-1} x_q x_q^\top,
	\qquad
	x_q =
	\begin{pmatrix}
		\cos(\lambda_k q)\\[1mm]
		\sin(\lambda_k q)
	\end{pmatrix},
	\qquad
	\lambda_k = \frac{2\pi k}{n}.
	\]
	Then
	\[
	Q_n
	=
	\frac{1}{n}\sum_{q=0}^{n-1}
	\begin{pmatrix}
		\cos^2(\lambda_k q) & \cos(\lambda_k q)\sin(\lambda_k q)\\[1mm]
		\cos(\lambda_k q)\sin(\lambda_k q) & \sin^2(\lambda_k q)
	\end{pmatrix}.
	\]
	
	Using the trigonometric identities
	\[
	\cos^2 t = \frac{1+\cos(2t)}{2},
	\qquad
	\sin^2 t = \frac{1-\cos(2t)}{2},
	\qquad
	\cos t \sin t = \frac{\sin(2t)}{2},
	\]
	we obtain
	\[
	Q_n
	=
	\frac12 I_2
	+
	\frac{1}{2n}
	\sum_{q=0}^{n-1}
	\begin{pmatrix}
		\cos(2\lambda_k q) & \sin(2\lambda_k q)\\[1mm]
		\sin(2\lambda_k q) & -\cos(2\lambda_k q)
	\end{pmatrix}.
	\]
	
	Since \(\lambda_k = \frac{2\pi k}{n}\), the sums
	\[
	\sum_{q=0}^{n-1} \cos(2\lambda_k q),
	\qquad
	\sum_{q=0}^{n-1} \sin(2\lambda_k q)
	\]
	are finite geometric sums of complex exponentials and satisfy
	\[
	\sum_{q=0}^{n-1} e^{i 2\lambda_k q}
	=
	\sum_{q=0}^{n-1} \left(e^{i 4\pi k / n}\right)^q
	=
	\frac{1 - e^{i 4\pi k}}{1 - e^{i 4\pi k / n}}
	= 0,
	\]
	for all integers \(k\not\equiv 0 \pmod{n/2}\).
	Hence,
	\[
	\sum_{q=0}^{n-1} \cos(2\lambda_k q) = 0,
	\qquad
	\sum_{q=0}^{n-1} \sin(2\lambda_k q) = 0.
	\]
	
	Therefore, the oscillatory terms vanish exactly, and we obtain
	\[
	Q_n = \frac12 I_2
	\quad\text{for all } n,
	\qquad\text{hence}\qquad
	Q = \lim_{n\to\infty} Q_n = \frac12 I_2.
	\]
	
	We now detail the computation of \(M_n(h)\).
	
	For a fixed lag \(h\in\mathbb Z\), define
	\[
	M_n(h)
	:=
	\frac{1}{n}\sum_{q=0}^{n-1-|h|}
	x_q x_{q+|h|}^\top,
	\qquad
	x_q=
	\begin{pmatrix}
		\cos(q\lambda_k)\\
		\sin(q\lambda_k)
	\end{pmatrix}.
	\]
	
	We first expand the matrix product
	\[
	x_q x_{q+h}^\top
	=
	\begin{pmatrix}
		\cos(q\lambda_k)\cos((q+h)\lambda_k)
		&
		\cos(q\lambda_k)\sin((q+h)\lambda_k)
		\\[1mm]
		\sin(q\lambda_k)\cos((q+h)\lambda_k)
		&
		\sin(q\lambda_k)\sin((q+h)\lambda_k)
	\end{pmatrix}.
	\]
	
	Using the standard product-to-sum trigonometric identities
	\[
	\begin{aligned}
		\cos a \cos b &= \tfrac12\bigl[\cos(a-b)+\cos(a+b)\bigr],\\
		\sin a \sin b &= \tfrac12\bigl[\cos(a-b)-\cos(a+b)\bigr],\\
		\cos a \sin b &= \tfrac12\bigl[\sin(b+a)+\sin(b-a)\bigr],\\
		\sin a \cos b &= \tfrac12\bigl[\sin(a+b)+\sin(a-b)\bigr],
	\end{aligned}
	\]
	with \(a=q\lambda_k\) and \(b=(q+h)\lambda_k\), we obtain
	\[
	x_q x_{q+h}^\top
	=
	\frac12
	\begin{pmatrix}
		\cos(h\lambda_k)+\cos((2q+h)\lambda_k)
		&
		\sin(h\lambda_k)+\sin((2q+h)\lambda_k)
		\\[1mm]
		-\sin(h\lambda_k)+\sin((2q+h)\lambda_k)
		&
		\cos(h\lambda_k)-\cos((2q+h)\lambda_k)
	\end{pmatrix}.
	\]
	
	Summing over \(q\) and dividing by \(n\), we get
	\[
	\frac{1}{n}\sum_{q=0}^{n-1-|h|} x_q x_{q+h}^\top
	=
	\frac12
	\begin{pmatrix}
		\cos(h\lambda_k) & \sin(h\lambda_k)\\[1mm]
		-\sin(h\lambda_k) & \cos(h\lambda_k)
	\end{pmatrix}
	+ R_n(h),
	\]
	where the remainder matrix \(R_n(h)\) is given by
	\[
	R_n(h)
	=
	\frac{1}{2n}\sum_{q=0}^{n-1-|h|}
	\begin{pmatrix}
		\cos((2q+h)\lambda_k) & \sin((2q+h)\lambda_k)\\[1mm]
		\sin((2q+h)\lambda_k) & -\cos((2q+h)\lambda_k)
	\end{pmatrix}.
	\]
	
	We now show that \(R_n(h)=o(1)\).
	Each entry of \(R_n(h)\) is a normalized sum of the form
	\[
	\frac{1}{n}\sum_{q=0}^{n-1} e^{i(2q+h)\lambda_k}
	=
	\frac{e^{ih\lambda_k}}{n}
	\sum_{q=0}^{n-1} \bigl(e^{i2\lambda_k}\bigr)^q.
	\]
	This is a geometric sum with ratio \(e^{i2\lambda_k}\).
	Under the non-degeneracy condition
	\[
	2\lambda_k \notin 2\pi\mathbb Z
	\quad
	\text{(equivalently \(k\not\equiv 0 \!\!\!\pmod{n/2}\))},
	\]
	we have \(e^{i2\lambda_k}\neq 1\), and therefore
	\[
	\sum_{q=0}^{n-1} \bigl(e^{i2\lambda_k}\bigr)^q
	=
	\frac{1-e^{i2n\lambda_k}}{1-e^{i2\lambda_k}}
	=
	O(1).
	\]
	Hence,
	\[
	\frac{1}{n}\sum_{q=0}^{n-1} e^{i(2q+h)\lambda_k}
	= O\!\left(\frac{1}{n}\right),
	\]
	and the same bound holds for the real and imaginary parts.
	Consequently,
	\[
	R_n(h)=O\!\left(\frac{1}{n}\right)=o(1).
	\]
	
	We conclude that
	\[
	\frac{1}{n}\sum_{q=0}^{n-1-|h|} x_q x_{q+h}^\top
	=
	\frac12
	\begin{pmatrix}
		\cos(h\lambda_k) & \sin(h\lambda_k)\\[1mm]
		-\sin(h\lambda_k) & \cos(h\lambda_k)
	\end{pmatrix}
	+ o(1),
	\]
	and therefore
	\[
	\lim_{n\to\infty} M_n(h)
	=
	\frac12 R(\lambda_k h).
	\]
	
	Under Assumption (A3) we have
	\(\gamma_\eta(h)=\mathrm{Cov}(\eta_0,\eta_h)\), we define
	\[
	\Gamma_h
	=
	\gamma_\eta(h)\, M(h)
	=
	\frac12\,\gamma_\eta(h)\,R(\lambda_k h),
	\]
	and the long-run covariance matrix is
	\[
	B
	=
	\sum_{h=-\infty}^{\infty} \Gamma_h
	=
	\frac12
	\sum_{h=-\infty}^{\infty}
	\gamma_\eta(h)\, R(\lambda_k h).
	\]
	
	Since \(A=2 f_\varepsilon(0)Q\) and \(Q=\tfrac12 I_2\), we have \(A=f_\varepsilon(0)I_2\). Therefore
	\[
	V = A^{-1} B A^{-1}
	= \frac{1}{f_\varepsilon(0)^2}\,B
	= \frac{1}{2 f_\varepsilon(0)^2}\sum_{h=-\infty}^{\infty}\gamma_\eta(h)\,R(\lambda_k h).
	\]
	This concludes the proof.
	\qed\\
	
	The limiting distribution of the LAD estimator established above constitutes a key intermediate result, as it provides the probabilistic basis for deriving the asymptotic behavior of the NKK periodogram, which is stated in the following theorem.\\
	
	\begin{theorem}
		Suppose that Assumptions (A1)--(A4) hold.
		Let $\lambda_k\in(0,\pi)$ be a  Fourier frequency.
		Define the NKK periodogram by
		\[
		N^{(J)}_k
		=
		\frac{n}{8\pi}\,
		\big\|\widehat{\beta}_{n}(\lambda_k)\big\|^2,
		\]
		where $\widehat{\beta}_{n}(\lambda_k)$ is the LAD estimator associated with
		the harmonic regressors
		\[
		x_q(\lambda_k)
		=
		\begin{pmatrix}
			\cos(q\lambda_k)\\[1mm]
			\sin(q\lambda_k)
		\end{pmatrix}.
		\]
		Then, as $n\to\infty$,
		\[
		N^{(J)}_k
		\ \xrightarrow{d}\
		\frac{1}{8\pi}\,
		Z^\top Z,
		\]
		where $Z$ is a centered bivariate Gaussian vector with covariance matrix
		\[
		\Sigma
		=
		\frac{1}{2 f_\varepsilon(0)^2}
		\sum_{h=-\infty}^{\infty}
		\gamma_\eta(h)\,
		R(\lambda_k h),
		\qquad
		\gamma_\eta(h)=\mathrm{Cov}(\eta_0,\eta_h),
		\]
		and
		\[
		R(\lambda_k h)=
		\begin{pmatrix}
			\cos\lambda_k h & \sin\lambda_k h\\[1mm]
			-\sin\lambda_k h & \cos\lambda_k h
		\end{pmatrix}.
		\]
		Equivalently, the limit distribution can be written as
		\[
		N^{(J)}_k
		\ \xrightarrow{d}\
		\frac{1}{8\pi}
		\sum_{j=1}^{2} \lambda_j\,\chi^2_{1,j},
		\]
		where $\lambda_1,\lambda_2$ denote the eigenvalues of $\Sigma$ and
		$\chi^2_{1,1},\chi^2_{1,2}$ are independent chi-square random variables
		with one degree of freedom.
	\end{theorem}
	\paragraph{Proof:}
	
	Recall that the NKK periodogram is defined by
	\[
	N^{(J)}_k
	=
	\frac{n}{8\pi}\,
	\big\|\widehat{\beta}_{n}(\lambda_k)\big\|^2,
	\]
	where $\widehat{\beta}_{n}(\lambda_k)$ denotes the LAD estimator associated
	with the harmonic regressors
	\[
	x_q(\lambda_k)
	=
	\begin{pmatrix}
		\cos(q\lambda_k)\\[1mm]
		\sin(q\lambda_k)
	\end{pmatrix}.
	\]
	
	\noindent From Theorem \ref{th1}, under Assumptions (A1)--(A4), we have
	\[
	\sqrt{n}\big(\widehat{\beta}_{n}(\lambda_k)-\beta_0(\lambda_k)\big)
	\ \xrightarrow{d}\
	Z,
	\qquad
	Z\sim \mathcal N(0,\Sigma),
	\]
	where the asymptotic covariance matrix is given by
	\[
	\Sigma
	=
	\frac{1}{2 f_\varepsilon(0)^2}
	\sum_{h=-\infty}^{\infty}
	\gamma_\eta(h)\,
	R(\lambda_k h),
	\]
	with $\gamma_\eta(h)=\mathrm{Cov}(\eta_0,\eta_h)$ and
	\[
	R(\lambda_k h)
	=
	\begin{pmatrix}
		\cos\lambda_k h & \sin\lambda_k h\\[1mm]
		-\sin\lambda_k h & \cos\lambda_k h
	\end{pmatrix}.
	\]
	
	Under the null hypothesis of no periodic component at frequency $\lambda_k$,
	we have $\beta_0(\lambda_k)=0$, and therefore
	\[
	\sqrt{n}\,\widehat{\beta}_{n}(\lambda_k)
	\ \xrightarrow{d}\
	Z.
	\]
	
	\noindent
	
	By squaring the norm and multiplying by $n$, we obtain
	\[
	n\,\big\|\widehat{\beta}_{n}(\lambda_k)\big\|^2
	=
	\big\|\sqrt{n}\,\widehat{\beta}_{n}(\lambda_k)\big\|^2.
	\]
	Since the mapping $x\mapsto \|x\|^2$ is continuous on $\mathbb R^2$,
	the continuous mapping theorem yields
	\[
	n\,\big\|\widehat{\beta}_{n}(\lambda_k)\big\|^2
	\ \xrightarrow{d}\
	Z^\top Z.
	\]
	
	Multiplying both sides by the deterministic constant $(8\pi)^{-1}$ and
	applying Slutsky’s theorem, we obtain
	\[
	N^{(J)}_k
	=
	\frac{n}{8\pi}\,
	\big\|\widehat{\beta}_{n}(\lambda_k)\big\|^2
	\ \xrightarrow{d}\
	\frac{1}{8\pi}\, Z^\top Z.
	\]
	
	Let $\lambda_1,\lambda_2$ denote the eigenvalues of the positive definite
	matrix $\Sigma$. Since $Z\sim\mathcal N(0,\Sigma)$, the quadratic form
	$Z^\top Z$ admits the spectral representation
	\[
	Z^\top Z
	\ \overset{d}{=}\ 
	\lambda_1\,\chi^2_{1,1}+\lambda_2\,\chi^2_{1,2},
	\]
	where $\chi^2_{1,1}$ and $\chi^2_{1,2}$ are independent chi-square random
	variables with one degree of freedom.
	
	Consequently,
	\[
	N^{(J)}_k
	\ \xrightarrow{d}\
	\frac{1}{8\pi}
	\sum_{j=1}^{2} \lambda_j\,\chi^2_{1,j}.
	\]
	
	This completes the proof.
	\hfill$\square$\\
		
	Having established the asymptotic distribution of the NKK periodogram under suitable regularity conditions, we now turn to a Monte Carlo study in order to assess the finite-sample behavior of the statistic and to illustrate the relevance of the theoretical results in practice.
	
	\section{Simulation Study}\label{sec_04}
    \noindent
    In this section, we empirically evaluate the asymptotic behavior of the NKK periodogram when the data are generated from long-memory processes exhibiting heavy tails. Contrary to the classical Gaussian framework, we exclusively consider innovations following a Student distribution, in order to analyze the robustness of the procedure in a non-Gaussian setting.\\
    
    We consider two classes of fractionally integrated autoregressive moving-average processes, ARFIMA (see \cite{Beran2013}).\\
    
    The first model corresponds to a pure fractional process, defined by
    \begin{equation*}
    	(1-L)^d Z_t = \varepsilon_t,
    \end{equation*}
    where $L$ denotes the lag operator, $d \in (0,\frac{1}{2})$ is the long-memory parameter, and $\{\varepsilon_t\}$ is a sequence of independent and identically distributed random variables following a Student distribution,
    \begin{equation*}
    	\varepsilon_t \sim t_\nu,
    \end{equation*}
    with $\nu = 5$, which guarantees finite variance while introducing heavy tails.
    
    The fractional $I(d)$ process is generated using the series representation
    \begin{equation*}
    	Z_t = \sum_{k=0}^{t-1} \frac{(d)_k}{k!} \, u_{t-k},
    	\qquad
    	(d)_k = d(d+1)\cdots(d+k-1),
    \end{equation*}
    where $\{u_t\}$ is Student white noise.
    
\begin{figure}[h!]
    	\centering
    	\includegraphics[width=0.9\textwidth]{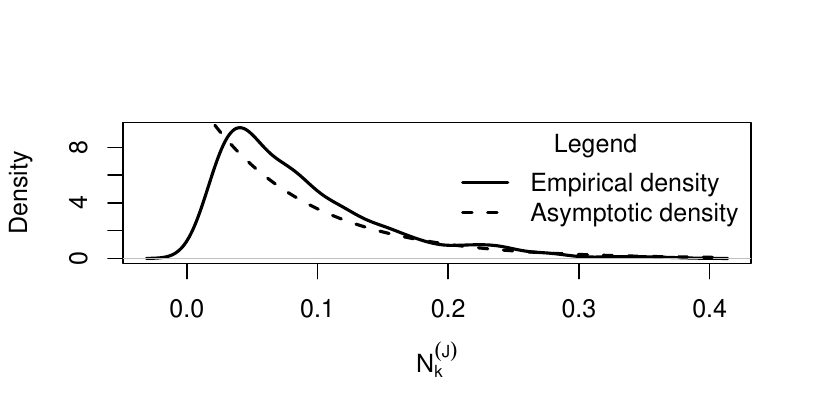}
    	\caption{Empirical and asymptotic densities of the NKK periodogram for an ARFIMA$(0,0.1,0)$ process with Student innovations ($\nu=5$).}
    	\label{fig:arfima0d0_1}
    \end{figure}

\begin{figure}[h!]
    	\centering
    	\includegraphics[width=0.9\textwidth]{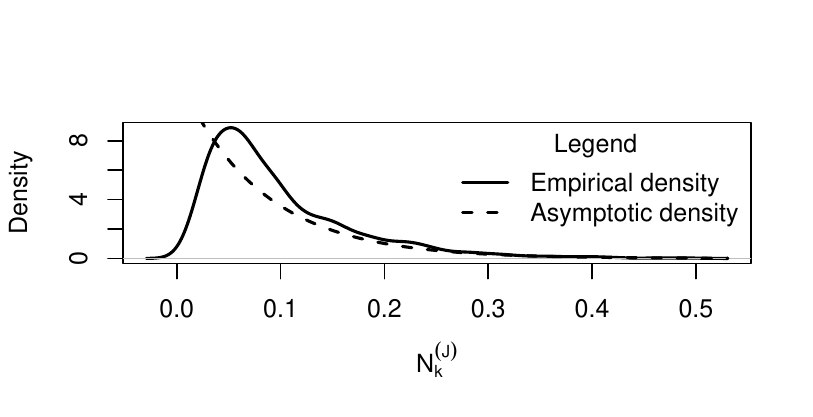}
    	\caption{Empirical and asymptotic densities of the NKK periodogram for an ARFIMA$(0,0.3,0)$ process with Student innovations ($\nu=5$).}
    	\label{fig:arfima0d0_2}
    \end{figure}

    The simulations are carried out for $d = 0.1$ and $d = 0.3$ (see Figures~\ref{fig:arfima0d0_1} and~\ref{fig:arfima0d0_2}, respectively), with a sample size $n = 1024$.\\
    
    The second model introduces short-term dynamics and is defined by
    \begin{equation*}
    	(1 - \phi L)(1-L)^d Z_t = (1 + \theta L)\varepsilon_t,
    \end{equation*}
    where $\phi$ and $\theta$ are the autoregressive and moving-average parameters, respectively.
    
\begin{figure}[h!]
    	\centering
    	\includegraphics[width=0.9\textwidth]{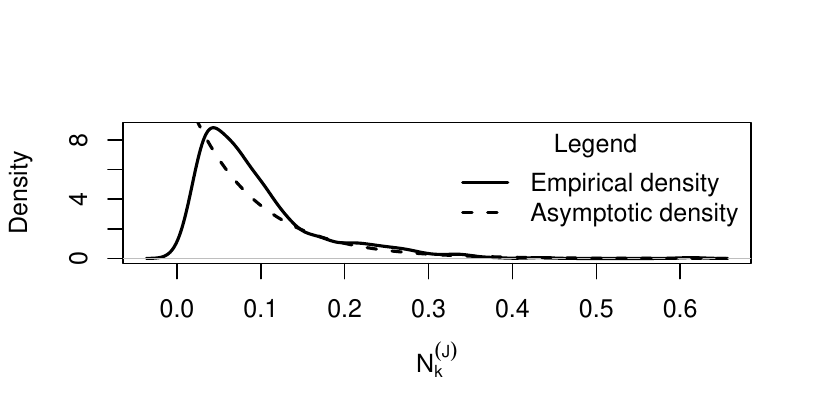}
    	\caption{Empirical and asymptotic densities of the NKK periodogram for an ARFIMA$(1,0.1,1)$ process with Student innovations ($\nu=5$).}
    	\label{fig:arfima1d1_1}
    \end{figure}

\begin{figure}[h!]
    	\centering
    	\includegraphics[width=0.9\textwidth]{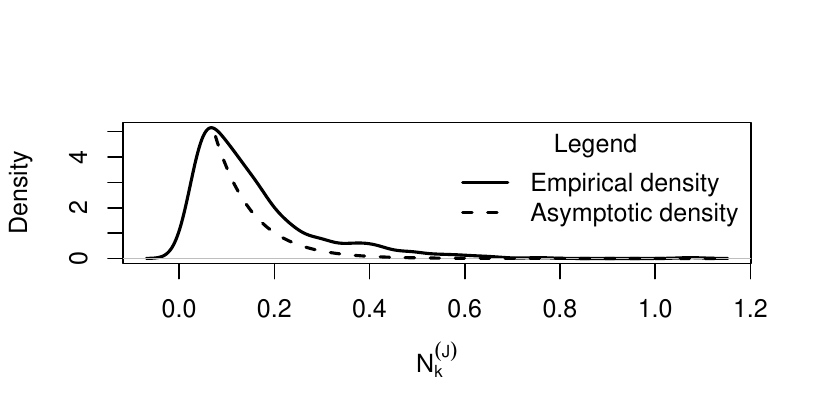}
    	\caption{Empirical and asymptotic densities of the NKK periodogram for an ARFIMA$(1,0.3,1)$ process with Student innovations ($\nu=5$).}
    	\label{fig:arfima1d1_2}
    \end{figure}

    The parameter configurations considered in the simulations are
    \[
    (\phi, \theta, d, \nu) \in \{(0.3, -0.2, 0.1, 5), \ (0.3, -0.2, 0.3, 5)\},
    \]
    with a sample size $n = 2048$, see Figures~\ref{fig:arfima1d1_1} and~\ref{fig:arfima1d1_2}, respectively.
    The innovations $\varepsilon_t$ also follow a Student distribution with $\nu = 5$ degrees of freedom.\\
    
    For each simulated trajectory, we apply the maximal overlap discrete wavelet transform (MODWT) using the \texttt{LA8} wavelet (see \cite{daubechi}), which is well suited for the analysis of long-range dependence.\\
    
    At each scale, the NKK periodogram is constructed using a LAD (Least Absolute Deviations) estimation of the Fourier coefficients. The final periodogram is obtained by averaging over all considered scales.\\
    
    For each configuration studied:
    \begin{itemize}
    	\item $1000$ independent replications are generated;
    	\item the statistic $N^{(J)}_k$ is computed at the lowest frequency ($k=1$);
    	\item the empirical distributions are compared with the theoretical asymptotic density given by
    	\begin{equation*}
    		\frac{1}{8\pi}\chi^2(2).
    	\end{equation*}
    \end{itemize}
    
    The results are presented through a comparison between kernel-based empirical density estimates of the NKK periodogram and the corresponding theoretical asymptotic density.\\
    
    The resulting density curves show a good agreement between the empirical behavior of the NKK periodogram and its theoretical asymptotic distribution, even when the innovations follow a heavy-tailed Student distribution (see Figure \ref{fig:arfima0d0_1} to Figure \ref{fig:arfima1d1_2}).
    
    In the case of ARFIMA$(0,d,0)$ processes (see Figure \ref{fig:arfima0d0_1} and Figure \ref{fig:arfima0d0_2}), the agreement is particularly satisfactory for moderate sample sizes. Introducing an ARMA structure in the ARFIMA$(1,d,1)$ case does not significantly alter the asymptotic behavior of the statistic (see Figure \ref{fig:arfima1d1_1} and Figure \ref{fig:arfima1d1_2}), although slight discrepancies may appear in the tails of the distribution, as expected in the presence of non-Gaussian innovations.
    
    Overall, these results confirm the robustness of the NKK periodogram with respect to heavy tails, the presence of short-term dependence, and variations in the long-memory parameter $d$.

    The simulation results provide strong empirical support for the theoretical findings and motivate the concluding discussion presented in the next section.
    \section{Conclusion}
    \label{concl}
    This paper has investigated the asymptotic behavior of a wavelet-based NKK periodogram constructed from least absolute deviation (LAD) harmonic regressions at a fixed scale. By exploiting the multiresolution properties of wavelet coefficients and a regression-based representation in the frequency domain, we established a non-degenerate limiting distribution for the NKK periodogram under suitable regularity and dependence conditions. The analysis highlights the role of scale-wise decorrelation induced by the wavelet transform and clarifies the impact of harmonic structure and frequency selection on the asymptotic behavior of the periodogram.
    
    The derived limit theorem provides a rigorous theoretical foundation for the use of robust, wavelet-based periodograms in the analysis of long-memory time series. In contrast to classical log-periodogram approaches, the NKK periodogram studied here exhibits well-defined asymptotic properties under heavy-tailed innovations and long-range dependence, while remaining compatible with a regression-based construction.
    
    The Monte Carlo simulations further confirm the theoretical results, showing that the empirical distribution of the NKK periodogram closely matches its asymptotic limit, even in the presence of heavy-tailed innovations and short-term dependence. These findings support the robustness of the proposed approach and highlight its practical relevance for the spectral analysis of long-memory processes.
    
    \bibliographystyle{plain}
	\end{document}